# Two-Step Dual-Wavelength Flash Raman Mapping Method for Measuring Thermophysical Properties of Supported Two-Dimensional Nanomaterials

Aoran Fan [a], Haidong Wang [a], Weigang Ma [a] and Xing Zhang [a,*]

[a] Key Laboratory for Thermal Science and Power Engineering of Ministry of Education,

Department of Engineering Mechanics, Tsinghua University, Beijing, 100084, China

**\*Contact:** Professor Xing Zhang, Key Laboratory for Thermal Science and Power Engineering of

Ministry of Education, Department of Engineering Mechanics, Tsinghua University, Beijing,

100084, China. Email: x-zhang@mail.tsinghua.edu.cn. Tel: 86-010-62772668.

*ABSTRACT*

To determine the thermophysical properties of supported two-dimensional nanomaterials, this paper developed a two-step dual-wavelength flash Raman mapping method. The thermal conductivity of the supported two-dimensional nanomaterial and the thermal contact conductance between the sample and the substrate can be determined by steady-state step. And then the thermal diffusivity of the sample can be characterized by the transient step. Two models are also proposed in this paper. When the substrate temperature rises obviously, a full model considered the substrate temperature distribution is used to decouple the thermal diffusivity, thermal conductivity, and the thermal contact conductance. When the maximum substrate temperature rise is less than 20% of the maximum sample temperature rise, the temperature distribution and variation of the substrate is assumed proportionate to the sample, and then a simplified model can be used to analyse the thermal diffusivity and the other parameters. For the thermal diffusivity, the system error caused by simplifying assumptions is less than 1%, when a dimensionless parameter related to the thermal contact conductance and the thermal conductivity less than 1 and the maximum temperature rise of the substrate less than 15% of that of the supported sample.

*INTRODUCTION*

Due to the superconductivity [1, 2], pyroelectricity [3], and other novel properties [4, 5], the two-dimensional nanomaterials have been the focus of recent research and could be the basement of the next-generation electronic productions [6-8]. Thus, the characterization of the thermophysical properties of nanomaterials is of great importance. However, in practical electronic devices, nanomaterials are usually supported on a substrate, and the interfacial electron or phonon scattering will also influence the thermophysical properties of supported nanomaterials compared with that of suspended nanomaterials [9-12]. Therefore, the in-situ measurement of the thermophysical properties of the supported nanomaterial is essential.

There are two significant methods on the measurement of the thermophysical properties of supported two-dimensional nanomaterials: the contact method [13-18] and the non-contact method [11, 12, 19-21]. For the contact method, researchers have used it to determine the thermal conductivity of supported graphene, and measured the thermal contact conductance between the graphene and the silicon dioxide substrate. However, the influence of electrical/thermal contact resistance between the electrode and the sample is hard to be eliminated, and the temperature rise of the substrate cannot be measured directly, which would increase the measurement uncertainty.

For non-contact method, the thermo-domain thermal reflection (TDTR) method [19-21] has been widely used to determine the thermophysical properties of the supported nanofilm, especially metal nanofilm. However, in the TDTR method, the metal sensor on the sample surface should be more than 10 nm, which makes it not an ideal method for ultrathin nanofilm. As another non-contact method, the Raman method seems as the most promising method to study supported nanomaterial. It can determine the temperature variation of the monatomic film directly, such as graphene, and can measure the temperature variation of the substrate simultaneously. However, for the supported nanofilm, the laser absorptivity is hard to measure, which will significantly affect the measurement accuracy of the most used steady-state Raman method [4, 11, 22]. Some methods [23, 24] have been developed to directly measure the optical absorptivity by combining electrical and optical measurement, however, the methods are still limited by complex nanofabrication.

The transient Raman method [12, 25-31] can eliminate the laser absorption influence by normalization, and this method has been developed to determine the thermal diffusivity of the suspended one-dimensional, two-dimensional nanomaterials, the supported van der waals heterostructures, and the thermal contact resistance between two carbon fibers. However, due to the limitation of temporal resolution and spatial resolution of previous measurement methods, the measurement sensitivity of the supported two-dimensional nanomaterial is still not satisfactory, and the influence of multi thermal parameters have not decoupled directly in the model yet, which will also increase the uncertainty.

In our recent work, a transient dual-wavelength flash Raman (DFR) mapping method [32-35] with 100 ps temporal resolution and 50 nm spatial resolution was proposed, and was further developed to determine the thermal diffusivity of suspended one-dimensional [35], two-dimensional [33], and anisotropic nanomaterials [34]. However, this method has not been applied to supported two-dimensional nanomaterials yet. Because for suspended nanomaterials, the thermal diffusivity is the only one effect factor of the non-dimensional transient heat conduction proceeding. But for the supported nanomaterials, the transient heat conduction proceeding can be also influenced by the thermal conductance between the sample and the substrate. More information is needed for the thermal diffusivity measurement of the supported sample.

Therefore, based on the DFR mapping method, this paper presents a two-step DFR mapping method which further considered the steady-state heat conduction proceeding. The heating laser and the probing laser have different wavelength to make the Raman spectra excited by probing laser can be distinguished. In practice, the wavelength of the probing laser is always longer than the heating laser. In the steady-state step, by changing the center distance between the continuous heating laser spot and the continuous probing laser spot, the temperature distribution of the sample and the substrate can be measured by their Raman peak shift, and then the thermal conductivity of the supported sample and thermal contact conductance between the sample and the substrate can be analyzed. Through the normalization approach to analyze the temperature difference between the supported sample and the substrate, this steady-state measurement can directly characterize the influence of laser absorptivity, which is significantly distinguished from the previous steady-state Raman methods [4, 11, 22]. In the transient step, the heating laser and probing laser are pulsed, and then the time delay between the two pulsed lasers is changed to determine the temperature variation of the sample and the substrate, and then the thermal diffusivity of the sample can be determined. Therefore, through this method, the influence of multi thermal parameters can be decoupled with high sensitivity.

Meanwhile, due to the small thermal capacity of nanofilm, the temperature of the substrate is always obviously smaller than it of the sample. When the absolute uncertainty is certain, the relative temperature measurement uncertainty of the substrate would be significantly larger than it of the sample, which makes the measurement of the substrate temperature distribution hard and inaccurate. In the previous methods, the small temperature rise of the substrate is always ignored [11-13]. This paper provides a new simplified model which just needs to measure the maximum temperature rise of the substrate. The error analysis shows that, when the substrate temperature rise is less than 15% of the sample temperature rise and a dimensionless parameter CC less than 1, the system error of the thermal diffusivity caused by simplifying assumptions is less than 1%, while the system error caused by ignoring substrate temperature rise can reach to 5.6% in this case. The simplified model can greatly decrease the system error compared with the model ignored substrate temperature rise. And compared with the full model, this model can simplify the measurement and enhance the feasibility of the method.

## FULL MODEL AND SENSITIVITY ANALYSIS

### *Physical model and solution*

As shown in Figure 1, a periodic pulse Gaussian laser beam was used to heat the two-dimensional nanomaterial sample and the substrate, and another laser pulse with different wavelength and negligible heat effect is used as a probe to simultaneously measure the temperature variations of the sample and the substrate. The heating pulse width is $t_h$ and the heating pulse interval, $t_c$, is long enough for the sample temperature to recover to the ambient temperature. When the wavelength of the heating laser differs from that of the probing laser, the Raman spectra excited by the heating pulse and the probing pulse can be distinguished, and the sample temperature variations can be determined from the Raman peak shifts excited by the probing pulse. The temperature variations are the average temperature variations during the pulse width of the probing pulse, $t_p$, with the temperature variations during the heating and cooling periods measured by changing the time delay between the heating and probing pulse, $t_d$. In addition, the position of the probing laser spot can be moved by a scanning mirror. With a fixed heating laser spot, the temperature variation at arbitrary time delay and the optional position can be measured. When the thermophysical properties of the substrate are known, through the steady-state measurement and the transient measurement, the contact thermal conductance $g$ between the substrate and the nanomaterial, the thermal diffusivity $\alpha_n$, and thermal conductivity $\lambda_n$ of the supported two-dimensional nanomaterial can be determined.

There are two steps to this method. The first one is the steady-state step. By changing the centre distance between the continuous heating laser spot and the continuous probing laser spot, the temperature distribution of the sample and the substrate can be measured, and then the thermal conductivity and thermal contact conductance can be analysed. The interfacial heat flux is more than 5 orders larger than the heat loss of radiation and convection [11]. Thus, in a vacuum, the heat loss of the environment can be neglected, and the heat conduction in steady-state can be expressed as:

$$\frac{\partial^2 \theta_n^{st}(r)}{\partial r^2} + \frac{1}{r}\frac{\partial \theta_n^{st}(r)}{\partial r} + \frac{\eta_n q_h}{\lambda_n \delta}\exp\left(-r^2/r_h^2\right) - \frac{g}{\lambda_n \delta}\left(\theta_n^{st}(r) - \theta_{sub}^{st}(r,0)\right) = 0 \quad (1)$$

$$\frac{\partial^2 \theta_{sub}^{st}(r,z,t)}{\partial r^2} + \frac{1}{r}\frac{\partial \theta_{sub}^{st}(r,z,t)}{\partial r} + \frac{\partial^2 \theta_{sub}^{st}(r,z,t)}{\partial z^2} = 0$$
$$-\lambda_{sub}\frac{\partial \theta_{sub}^{st}(r,0)}{\partial z} = g \times \left(\theta_n^{st}(r) - \theta_{sub}^{st}(r,0)\right) + \eta_{sub} q_h \exp\left(-r^2/r_h^2\right) \quad (2)$$

where $\theta_n^{st}(r)$ is the temperature rise of the supported two-dimensional nanomaterial sample at radius $r$ in steady state, $\theta_{sub}^{st}(r,z)$ is the temperature rise of the substrate at radial position $r$ and axial position $z$ in steady state, $g$ is the contact thermal conductance between the sample and the substrate, $\eta_n$, $\lambda_n$ and $\delta$ are respectively the effective laser absorptivity, the thermal conductivity and the thickness of the sample, $\eta_{sub}$ and $\lambda_{sub}$ are respectively the effective laser absorptivity and the thermal conductivity of the substrate, $q_h$ is the laser power density of the heating pulse at the beam center, and $r_h$ is the laser spot radius of the heating pulse where the power density attenuates to $q_h/e$.

Due to the interaction effect between the supported nanomaterial and the substrate, the decoupling of the parameters will be complicated. To simply the parameter decoupling, the temperature difference between the supported two-dimensional nanomaterial sample and the substrate is considered as an independent variable. Then the heat conduction of the supported sample in steady-state can be expressed as:

$$\frac{\partial^2 \theta_n^{st}(r)}{\partial r^2} + \frac{1}{r}\frac{\partial \theta_n^{st}(r)}{\partial r} + \frac{\eta_n q_h}{\lambda_n \delta}\exp(-r^2/r_h^2) - \frac{g}{\lambda_n \delta}f^{st}(r) = 0 \qquad (3)$$

where $f^{st}(r) = \theta_n^{st}(r) - \theta_{sub}^{st}(r,0)$ is the temperature difference between the supported nanomaterial and the substrate in steady-state. By applying Hankel transform, $f^{st}(r)$ can be described as:

$$f^{st}(r) = \frac{\lambda_n \delta}{g}\left(\frac{\eta_n}{\lambda_n}\frac{q_h \exp(-r^2/r_h^2)}{\delta} - \sum_{m=1}^{\infty}\mu_m^2 \theta_n^{st,*}(\mu_m) K_0(\mu_m, r)\right) \qquad (4)$$

where the kernel $K_0(\mu_m, r)$ is

$$K_0(\mu_m, r) = \frac{\sqrt{2}}{R}\frac{J_0(\mu_m r)}{J_1(\mu_m R)} \quad (R \to \infty) \qquad (5)$$

where $J_0$ and $J_1$ are the zero-order and first-order Bessel functions of the first kind. The characteristic values, $\mu_m$, are $\mu_m = \beta_m/R$, $\beta_m$ are the positive roots of the first-order Bessel functions, $R$ is the dimensionless characteristic length. And the $\theta_n^{st,*}(\mu_m)$ is the Hankel transformation of $\theta_n^{st}(r)$

$$\theta_n^{st,*}(\mu_m) = \int_0^\infty \theta_n^{st}(r) K_0(\mu_m, r) r dr \tag{6}$$

Therefore, the parameter $\eta_n/\lambda_n$ can be determined by the relationship between the normalized temperature difference, $f^{st}(r)/f^{st}(0)$, and the temperature distribution of the sample, $\theta_n^{st}(r)$, then the parameter $\lambda_n/g$ can be then determined with the measured $\eta_n/\lambda_n$.

Correspondingly, the heat conduction of the substrate in steady state can be expressed as:

$$\frac{\partial^2 \theta_{sub}^{st}(r,z)}{\partial r^2} + \frac{1}{r}\frac{\partial \theta_{sub}^{st}(r,z)}{\partial r} + \frac{\partial^2 \theta_{sub}^{st}(r,z)}{\partial z^2} = 0$$
$$-\lambda_{sub}\frac{\partial \theta_{sub}^{st}(r,0)}{\partial z} = g \times f^{st}(r) + \eta_{sub} q_h \exp(-r^2/r_h^2) \tag{7}$$

where $\theta_{sub}^{st}(r,z)$ is the temperature distributions of the substrate in steady-state. By applying Hankel transform, $f^{st}(r)$ can be described as a function of the temperature of the substrate surface:

$$f^{st}(r) = \frac{\lambda_{sub}}{g}\left(\sum_{m=1}^{\infty} \mu_m \theta_{sub}^{st,*}(\mu_m, 0) K_0(\mu_m, r) - \frac{\eta_{sub}}{\lambda_{sub}} q_h \exp(-r^2/r_h^2)\right) \tag{8}$$

where $\theta_{sub}^{st,*}(\mu_m, 0)$ is the Hankel transformation of $\theta_{sub}^{st}(r,0)$,

$$\theta_{sub}^{st,*}(\mu_m, 0) = \int_0^\infty \theta_{sub}^{st}(r,0) K_0(\mu_m, r) r dr \tag{9}$$

Therefore, the parameter $\eta_{sub}/\lambda_{sub}$ can be determined by the relationship between the normalized temperature difference, $f^{st}(r)/f^{st}(0)$, and the temperature distribution of the substrate surface, $\theta_{sub}^{st}(r,0)$, and then $\lambda_{sub}/g$ can be then determined with the measured $\eta_{sub}/\lambda_{sub}$.

With the known thermal conductivity of the substrate, $\lambda_{sub}$, the contact thermal conductance $g$ can be characterized. And then the thermal conductivity, $\lambda_n$, and laser absorptivity, $\eta_n$, of the supported nanomaterial sample can be determined with the measured $\lambda_n/g$ and $\eta_n/\lambda_n$.

The second step of this method is the transient step, the heating laser and probing laser are pulsed, and then the time delay between the two pulsed lasers are changed to determine the temperature variation of the sample and the substrate, and then the thermal diffusivity of the sample can be determined. With the temperature difference between the sample and the substrate surface denoted as $f(r,t)=\theta_n(r,t)-\theta_{sub}(r,0,t)$, the heat conduction of the supported two-dimensional nanomaterial sample can be expressed as Eq. (10).

$$\frac{\partial^2 \theta_n(r,t)}{\partial r^2} + \frac{1}{r}\frac{\partial \theta_n(r,t)}{\partial r} + \frac{\Phi_n(r,t)}{\delta} = \frac{1}{\alpha_n}\frac{\partial \theta_n(r,t)}{\partial t}$$

$$\Phi_n(r,t) = \begin{cases} \frac{\eta_n q_h}{\lambda_n}\exp(-r^2/r_h^2) - \frac{g}{\lambda_n} \times f(r,t) & (t \leq t_h) \\ -\frac{g}{\lambda_n} \times f(r,t) & (t > t_h) \end{cases} \quad (10)$$

where $\theta_n(r, t)$ is the temperature rise of the supported two-dimensional nanomaterial sample at radius $r$ and time $t$, $\theta_{sub}(r, z, t)$ is the temperature rise of the substrate at position $(r, z)$ and time $t$, $\alpha_n$ is the thermal diffusivities of the sample.

The initial temperature rise of the supported nanomaterial is 0; and the boundary conditions at $r = 0$ are given by the symmetry conditions as

$$\frac{\partial \theta_n(0,t)}{\partial r} = 0 \quad (11)$$

The boundary conditions at infinity are

$$\theta_n(\infty,t) = 0 \quad (12)$$

The analytical solution for the temperature rises of the supported two-dimensional nanomaterial sample and the substrate are obtained by successively applying Hankel and Laplace transforms

$$\theta_n(r,t) = \sum_{m=1}^{\infty} \theta_n^*(\mu_m,t) K_0(\mu_m,r) \qquad (13)$$

where the $\theta_n^*(\mu_m,t)$ is the Hankel transformation of the temperature variation of the supported nanomaterial, which is obtained from the inverse Laplace transformation,

$$\theta_n^*(\mu_m,t) = \begin{cases} \begin{bmatrix} \dfrac{\eta_n q_h e^*(\mu_m)}{\mu_m^2 \lambda_n \delta}\left(1-\exp(-\alpha_n \mu_m^2 t)\right) \\ -\dfrac{\alpha_n g}{\lambda_n \delta}\int_0^t f^*(\mu_m,\tau)\exp(-\alpha_n \mu_m^2(t-\tau))d\tau \end{bmatrix} & (t \leq t_h) \\ \begin{bmatrix} \theta_n^*(\mu_m,t_h)\exp(-\alpha_n \mu_m^2 t) \\ -\dfrac{\alpha_n g}{\lambda_n \delta}\int_0^t f^*(\mu_m,\tau)\exp(-\alpha_n \mu_m^2(t-\tau))d\tau \end{bmatrix} & (t > t_h) \end{cases} \qquad (14)$$

where $e^*(\mu_m)$ and $f^*(\mu_m,t)$ are the Hankel transformation of $\exp(-r^2/r_h^2)$ and $f(r,t)$, respectively.

$$e^*(\mu_m) = \int_0^R \exp(-r^2/r_h^2) K_0(\mu_m,r) r dr \qquad (15)$$

$$f^*(\mu_m,t) = \int_0^R f(r,t) K_0(\mu_m,r) r dr \qquad (16)$$

According to the above solution, with the measured $\eta_n/\lambda_n$ and $g/\lambda_n$, the thermal diffusivity of the sample, $\alpha_n$, can be determined.

*Sensitivity analysis of full model*

Several cases were simulated to analyse the sensitivity of the two-step DFR mapping method. Figure 2 shows the sensitivity of the temperature difference $f^{st}(r)$ in a steady state with a certain temperature distribution of substrate surface $\theta_{sub}^{st}(r,0)$ calculated by a silicon substrate with 35 K temperature rise. For typical $\eta_{sub}/\lambda_{sub}$ and $\lambda_{sub}/g$ with ±10% variations, when the laser spot radius of the heating laser $r_h$ is 259 nm, the maximum sensitivities to the various $\eta_{sub}/\lambda_{sub}$ from $5\times10^{-7}$ m·K/W to $2\times10^{-6}$ m·K/W are respectively 3.2%, 5.3% and 6.1%, while the maximum sensitivities to the various $\lambda_{sub}/g$ is 10% when $r=0$.

Figure 3 shows the sensitivity of the temperature difference $f^{st}(r)$ in a steady state with a certain temperature distribution of the supported sample $\theta_n^{st}(r)$ calculated by a supported graphene with 50 K temperature rise. As shown in Figure 4(a), the normalized temperature difference is more sensitive to small parameter $\eta_n/\lambda_n$. When $\eta_n/\lambda_n = 5\times10^{-6}$ m·K/W, with $\eta_n/\lambda_n$ changes ±10%, the normalized temperature variation at $r = 500$ nm can reach to -0.10/+0.26. While the maximum sensitivities to the various $\lambda_n/g$ is also 10% when $r=0$.

With the determined thermal contact conductance $g$ and thermal conductivity $\lambda_n$, the thermal diffusivity can be determined by the normalized temperature rise curves of the supported sample as shown in Figure 4. In this case, the temperature difference $f(r,t)$ is assumed as 0.01 of the temperature rise of the corresponding suspended sample. The temperature variation curves are more sensitive in cooling proceeding. When thermal diffusivity changes ±10%, the maximum normalized temperature variation to the various $\alpha_n$ from $1\times10^{-3}$ m$^2$/s to $1\times10^{-5}$ m$^2$/s are respectively ±0.024, ±0.019 and ±0.023. Therefore, when the measurement uncertainty of normalized temperature is 0.01, the measurement uncertainty of the thermal diffusivity caused by transient temperature measurement is about ±5%.

## *SIMPLIFIED MODEL AND FEASIBILITY ANALYSIS*

### *Simplified model for supported two-dimensional nanomaterials*

In practice, the temperature rise of the substrate is always significantly lower than it of the supported two-dimensional nanomaterial, and the temperature distribution of the substrate would be hard to determine. Fortunately, in this case, the influence of the substrate temperature distribution on the supported sample would be also decreased. Thus, the temperature distribution of the substrate can be assumed as $\theta_{sub}^{st}(r,0) = \delta T_{ste} \times \theta_n^{st}(r)$, where $\delta T_{ste} = \theta_{sub}^{st}(0,0)/\theta_n^{st}(0)$. With the nondimensionalized techniques, the heat conduction equation in steady-state can be written as:

$$\frac{d^2 T_{ste}(x)}{dx^2} + \frac{1}{x}\frac{dT_{ste}(x)}{dx} - CC(1-\delta T_{ste})\times T_{ste}(x) + \exp(-x^2) = 0 \qquad (17)$$

where $x=r/r_h$ is the dimensionless coordinate with the characteristic length, $r_h$, $T_{ste} = \theta_n^{st}/\theta_0$ is the dimensionless temperature rises of the supported sample with $\theta_0 = \eta_n q_h r_h^2/\lambda_n \delta$ as the characteristic temperature rise, and $CC = g r_h^2/\lambda_n \delta$ is a dimensionless number which influence the heat transfer between the supported sample and the substrate. The analytical solution of the sample temperature rise $T_{st}$ can be obtained by applying Hankel transform, and which is given by

$$T_{ste}(x) = \sum_{n=1}^{\infty} \frac{E^*(\mu_m) K_0(\mu_m, x)}{\mu_m^2 + CC_{ste}} \quad (18)$$

where $E^*(\mu_m) = \int_0^R \exp(-x^2) K_0(\mu_m, x) x \, dx$, and $CC_{ste} = CC \times (1 - \delta T_{ste})$.

Analogously, in transient measurement, the temperature variation of the substrate can be assumed as $\theta_{sub}(r,0,t) = \delta T \times \theta_n(r,t)$, where $\delta T = \theta_{sub}(0,0,t_h) / \theta_n(0,t_h)$. Using the same characteristic length $r_h$, the same characteristic temperature $\theta_0$, and a characteristic time $t_0$, the nondimensionalized transient heat conduction of the sample can be described as:

$$\frac{\partial^2 T(x,\tau)}{\partial x^2} + \frac{1}{x} \frac{\partial T(x,\tau)}{\partial x} - CC(1 - \delta T) \times T(x,\tau) + \Phi(x,\tau) = \frac{1}{Fo} \frac{\partial T(x,\tau)}{\partial \tau}$$

$$\Phi(r,t) = \begin{cases} \exp(-x^2) & (\tau \leq \tau_h) \\ 0 & (\tau > \tau_h) \end{cases} \quad (19)$$

where $T = \theta_n / \theta_0$ is the transient dimensionless temperature rises of the supported sample with $\theta_0$ as the characteristic temperature rise, $\tau = t/t_0$, $\tau_h = t_h/t_0$, and $Fo = \alpha_n t_0 / r_h^2$.

By Hankel transform, the analytical solution of the sample temperature variation can be obtained as:

$$T(x,\tau) = \sum_{m=1}^{\infty} T^*(\mu_m, \tau) K_0(\mu_m, x) \quad (20)$$

where $T^*(\mu_m, \tau)$ is expressed as Eq. (21) with $CC_e = CC(1 - \delta T)$.

$$T^*(\mu_m, \tau) = \begin{cases} \dfrac{E^*(\mu_m)}{\mu_m^2 + CC_e} \left[1 - \exp\left(-Fo(\mu_m^2 + CC_e)\tau\right)\right] & (\tau \leq \tau_h) \\ T^*(\mu_m, \tau_h) \exp\left(-Fo(\mu_m^2 + CC_e)(\tau - \tau_h)\right) & (\tau > \tau_h) \end{cases} \quad (21)$$

Therefore, during the steady-state measurement, with the measured normalization temperature distribution of the supported sample and the steady-state temperature ratio $\delta T_{st}$, the dimensionless number $CC$ can be determined by Eq. (18). With the measured $CC$ and the transient temperature ratio $\delta T$, the Fourier number $Fo = \alpha_n t_0/r_h^2$ can be characterized by normalized temperature variation of the supported sample during heating and cooling proceeding. Due to the $t_0$ is a parameter chosen by the researcher to ensure the equation convergence, and $r_h$ is the laser radius determined by the heating laser wavelength and the numerical aperture of the optical, thus the thermal diffusivity $\alpha_n$ can be determined. With known density $\rho$ and specific heat $c_p$, the thermal conductivity $\lambda_n = \alpha_n \rho c_p$ can be further characterized. With known sample thickness $\delta$, the thermal contact conductance $g$ can be calculated by $g = CC \lambda_n \delta / r_h^2$.

*Sensitivity analysis of simplified model*

Figure 5(a) shows the normalized temperature distribution $\Theta^{st}$ in steady-state with various parameter $CC_{ste}$, and Figure 5(b) shows the measurement sensitivity $\Delta\Theta^{st}$, which is defined as $\Delta\Theta^{st} = |\Theta^{st}(CC_{ste}+20\%) - \Theta^{st}(CC_{ste}-20\%)|$. It can be observed that the sample with the larger parameter $CC_{ste}$ has a higher maximum sensitivity, while the sample with the smaller parameter $CC_{ste}$ has a larger sensitive region.

When choose $t_0 = l^2/\alpha_n$, $l$ is the characteristic length, the Fourier number $Fo$ will always be 1 for any sample. Figure 6 shows the transient normalized temperature variations $\Theta$ during transient measurement at different positions and with different parameter $CC_e$, where $x_{h-p}$ describes the dimensionless length between the centers of the heating laser spot and the probing laser spot. It can be observed that for smaller $CC_e$, the temperature variation curves would have higher sensitivity with a larger $x_{h-p}$, while the sensitivity difference would be no noticeable when $CC_e$ increase.

*Feasibility analysis of the simplified model*

The feasibility of the assumption used in the simplified model should be analysed. Figure 7 shows the system error of $CC = gr_h^2/\lambda_n\delta$ caused by the simplifying assumption with various $\delta T_{ste}$ and $CC$. It can be observed that when $\delta T_{ste}$ less than 0.05, the system error of $CC$ caused by simplifying assumption will less than 0.1%, while the system error caused by ignoring the substrate temperature rise will be almost 5%. The simplified model significantly decreases the system error compared with the model ignored the substrate temperature rise. For most measurement cases, the temperature ratio $\delta T_{ste}$ will less than 0.15 and $CC$ less than 1, and the system error of $CC$ no more than 1.1%, which shows the simplified model will be feasible for most steady-state measurements.

Figure 8 shows the system error of $\alpha_n$ caused by the simplifying assumption with various $\delta T$ and $CC$. With $CC$ increases, the thermal contact conductance $g$ increases, and the influence of substrate temperature variation on the sample temperature variation increases and makes the system error increase. Meanwhile, with $g$ increases, the temperature variation curves of the sample become more like the temperature variation curves of the substrate and which makes the system error decreases. The two opposite factors make an extremum of the system error in the thermal diffusivity $\alpha_n$ measurement. It can be observed that when $\delta T$ less than 0.15 and $CC$ less than 1, the system error of $CC$ caused by simplifying assumption will less than 1%. Correspondingly, as shown in Figure 9, with the same *δT* and *CC*, the system error caused by ignoring substrate temperature rise can reach to 5.6%. This further shows the advantages of the simplified model compared with the model ignored substrate temperature rise.

*CONCLUSIONS*

This paper presents a two-step DFR method for determining the thermal diffusivity, thermal conductivity, and contact thermal conductance of supported two-dimensional materials. In the first steady-state step, a continuous heating laser is used to heat the sample and the substrate to steady-state, another continuous probe laser, with a different wavelength and the neglectable heating effect, is used to determine the temperature distributions of the sample and the substrate. The thermal conductivity and the contact thermal conductance can be extracted by fitting the normalized temperature distribution of the sample and the substrate. In the second transient step, the pulsed heating laser is used to heat the sample and the substrate, and the probing pulse laser with various time delay is used to measure the temperature rises of the sample and the surface of the substrate simultaneously, then the thermal diffusivities $\alpha_n$ can be determined.

Furthermore, two models are proposed in this paper. When the substrate temperature rises obviously, a full model considered the substrate temperature distribution is used to decouple the thermal diffusivity, thermal conductivity, and the thermal contact conductance. When the measurement uncertainty of normalized temperature is 0.01, with a full model the measurement uncertainty of the various thermal diffusivity caused by transient temperature measurement is about

±5%. When the maximum substrate temperature rise is less than 20% of the maximum sample temperature rise, the temperature distribution and variation of the substrate is assumed proportionate to the sample, and then a simplified model can be used to analyze the thermal diffusivity and the other parameters. For the thermal diffusivity, the system error caused by simplifying assumptions is less than 1% when a dimensionless parameter *CC* less than 1 and the maximum temperature rise of the substrate less than 15% of that of the supported sample, while, with the same substrate temperature rise and *CC*, the system error caused by ignoring substrate temperature rise can reach to 5.6%. The simplified model can greatly decrease the system error compared with the model ignored substrate temperature rise. And compared with the full model, this model can simplify the measurement and enhance the feasibility of the method.

*NOMENCLATURE*

$c_p$     specific heat [J kg$^{-1}$ K$^{-1}$]

*CC*     dimensionless parameter $CC = gr_h^2/\lambda_n \delta$

DFR     dual-wavelength flash Raman

*Fo*     Fourier number

*f*     temperature difference between the supported nanomaterial and substrate [K]

*g*     thermal contact conductance [W m$^{-2}$ K$^{-1}$]

$J_0$     zero-order Bessel function

| | |
|---|---|
| $J_1$ | first-order Bessel function |
| $K$ | kernel function |
| $l$ | characteristic length [m] |
| $q$ | laser power density [W m$^{-2}$] |
| $r$ | cylindrical coordinate [m] |
| $R$ | dimensionless characteristic length |
| $r_{h\text{-}p}$ | distance between the centers of heating laser spot and probing laser spot [m] |
| $t$ | time [s] |
| $T$ | normalized temperature rise |
| $t_c$ | heating pulse interval [s] |
| $t_d$ | time delay between the heating and probing pulse [s] |
| TDTR | thermo-domain thermal reflection |
| $t_h$ | heating pulse width [s] |
| $t_p$ | probing pulse width [s] |
| $x$ | dimensionless coordinate with characteristic length |
| $x_{h\text{-}p}$ | dimensionless distance between the centers of heating laser spot and probing laser spot |
| $z$ | cylindrical coordinate [m] |

*Greek symbols*

| | | |
|---|---|---|
| $\alpha$ | | thermal diffusivity [m² s⁻¹] |
| $\beta_m$ | | positive roots of the first-order Bessel function |
| $\delta$ | | thickness of two-dimensional nanomaterial [m] |
| $\eta$ | | laser absorptivity |
| $\lambda$ | | thermal conductivity [W m⁻¹ K⁻¹] |
| $\theta$ | | temperature rise [K] |
| $\mu_m$ | | the ratio of characteristic roots of Bessel function to $R$ |
| $\tau$ | | characteristic time |
| $\tau_d$ | | dimensionless time delay between the heating and probing pulse |
| $\tau_h$ | | dimensionless heating pulse width |
| $\rho$ | | density [kg m⁻³] |
| $\Theta$ | | normalized temperature distribution |

*Subscripts*

| | |
|---|---|
| $e$ | effective value of transient state |
| $h$ | heating laser |
| $n$ | two-dimensional nanomaterial |
| $p$ | probing laser |
| $ste$ | effective value of steady state |
| *sub* | substrate |
| 0 | characteristic value |

*Superscripts*

| | |
|---|---|
| *st* | steady state |
| * | Hankel transformation |

## ACKNOWLEDGMENTS

This work was supported by the National Natural Science Foundation of China (Grant Nos. 51827807 and 51636002).

**LIST OF FIGURE CAPTIONS**

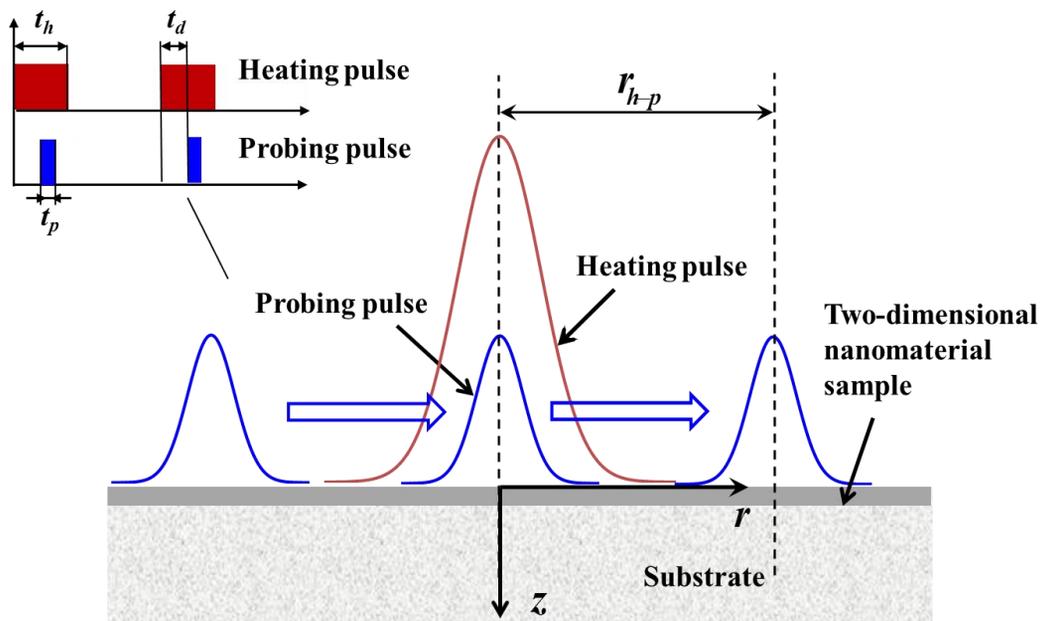

**Figure 1. Schematic of the supported two-dimensional nanomaterial sample and the substrate.**

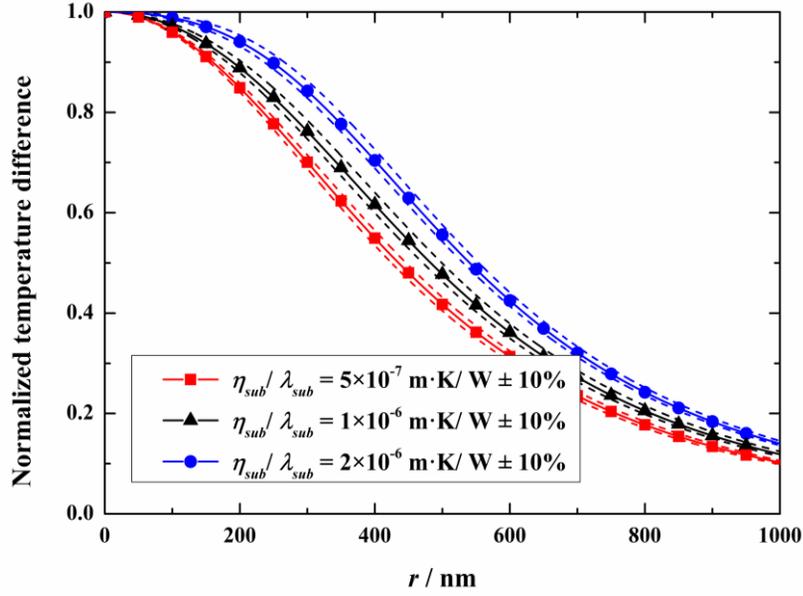

(a) Normalized temperature difference $f^{st}(r)/f^{st}(0)$ for various $\eta_{sub}/\lambda_{sub}$

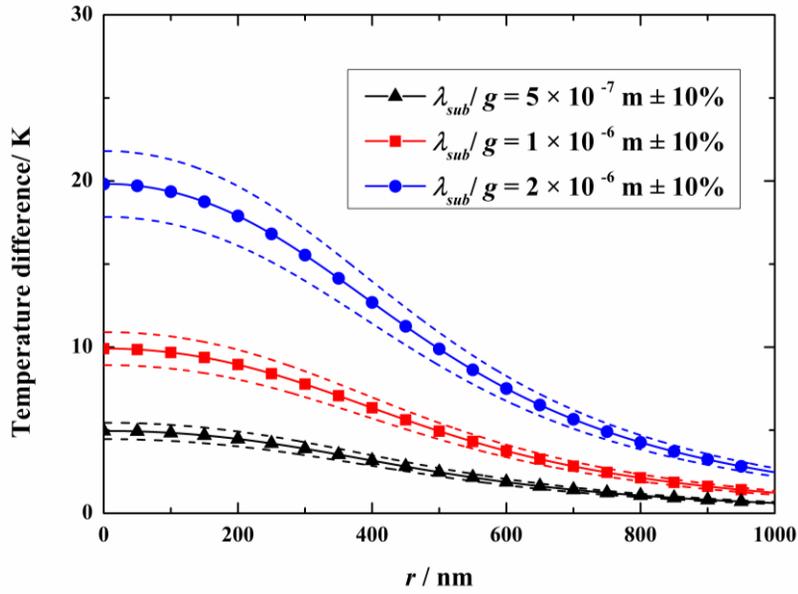

(b) Temperature difference $f^{st}(r)$ for various $\lambda_{sub}/g$ with a certain $\theta_{sub}^{st}(r,0)$

Figure 2. (a) Normalized temperature difference $f^{st}(r)/f^{st}(0)$ for various $\eta_{sub}/\lambda_{sub}$; (b) Temperature difference $f^{st}(r)$ for various $\lambda_{sub}/g$ with a certain $\theta_{sub}^{st}(r,0)$.

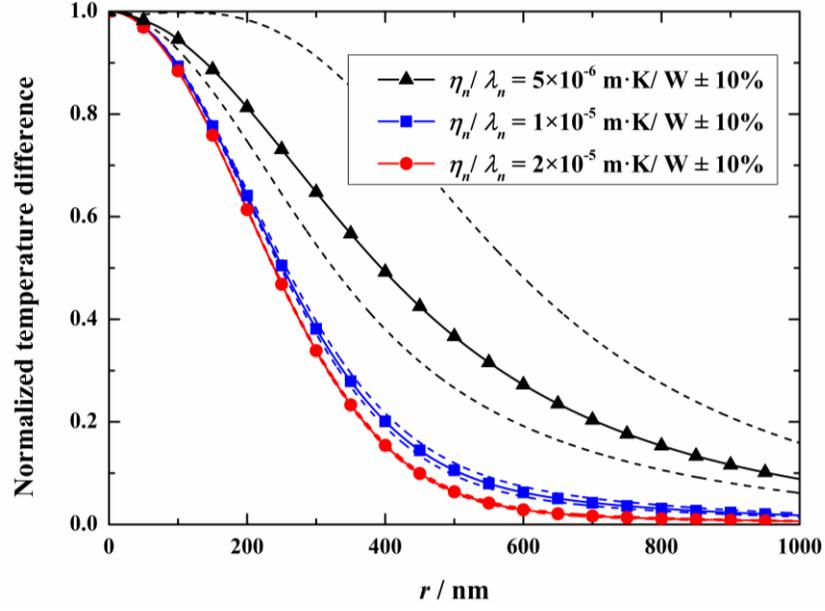

(a) Normalized temperature difference $f^{st}(r)/f^{st}(0)$ for various $\eta_n/\lambda_n$

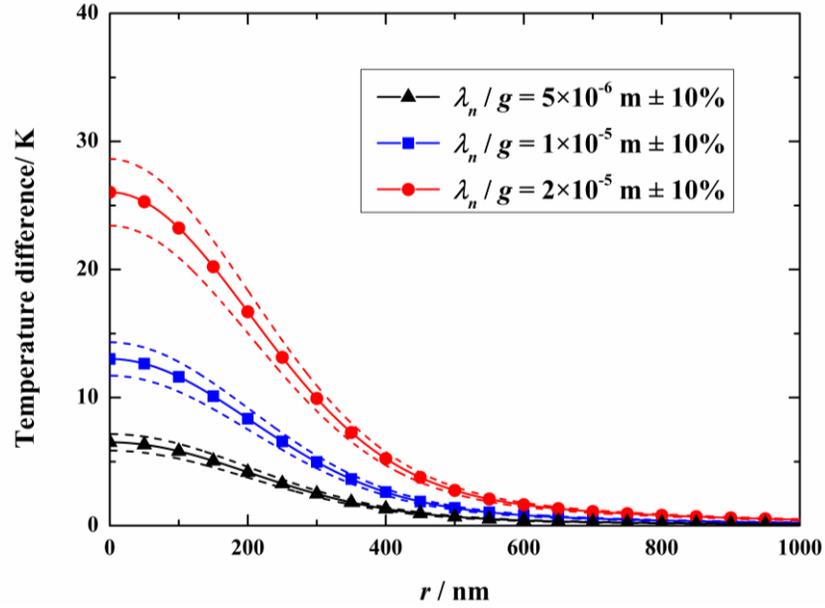

(b) Temperature difference for $f^{st}(r)$ various $\lambda_n/g$ with a certain $\theta_n^{st}(r)$

Figure 3. (a) Normalized temperature difference $f^{st}(r)/f^{st}(0)$ for various $\eta_n/\lambda_n$;

(b) Temperature difference for $f^{st}(r)$ various $\lambda_n/g$ with a certain $\theta_n^{st}(r)$.

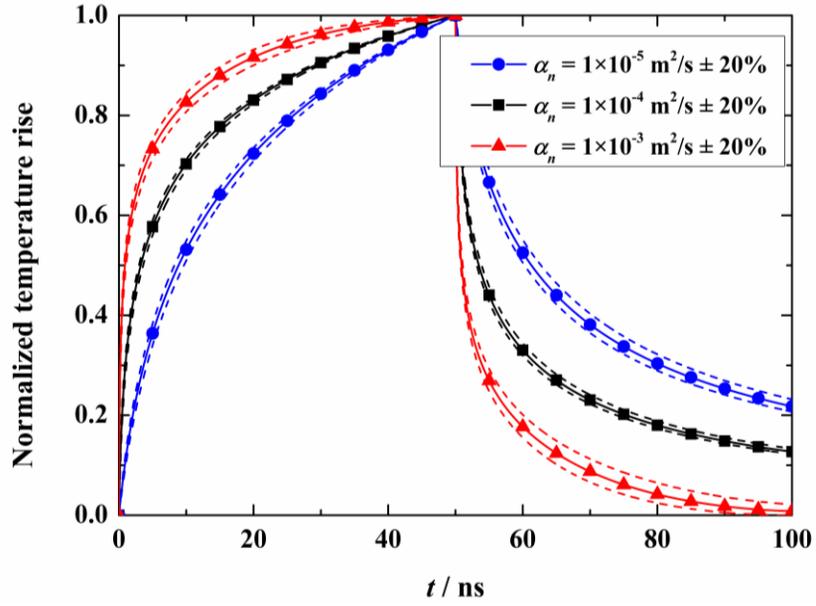

**Figure 4.** Normalized temperature rise curves of the supported sample for various $α_n$ ($g/λ_n=3 \times 10^4$ m$^{-1}$, heating pulse width $t_h$ = 50 ns, temperature difference is assumed as 0.01 of the temperature rise of the corresponding suspended sample).

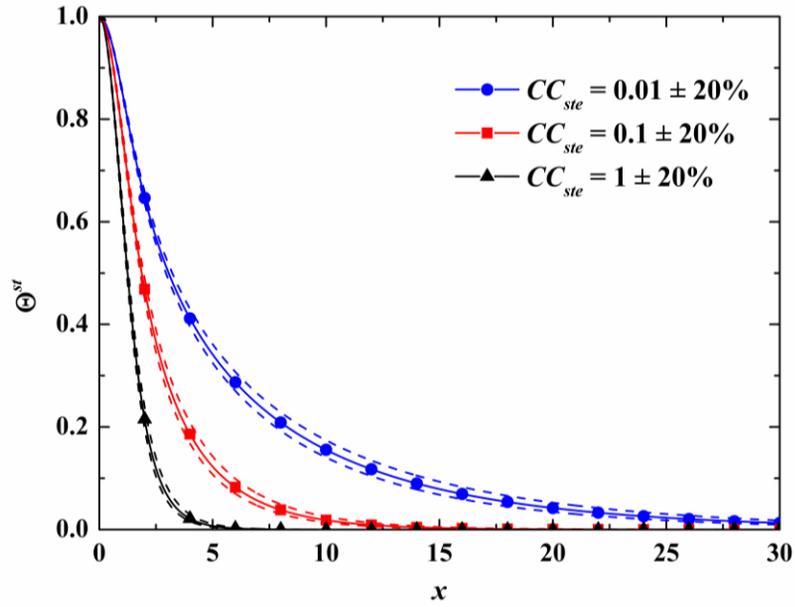

**(a) Normalized temperature distributions for various $CC_{ste}$**

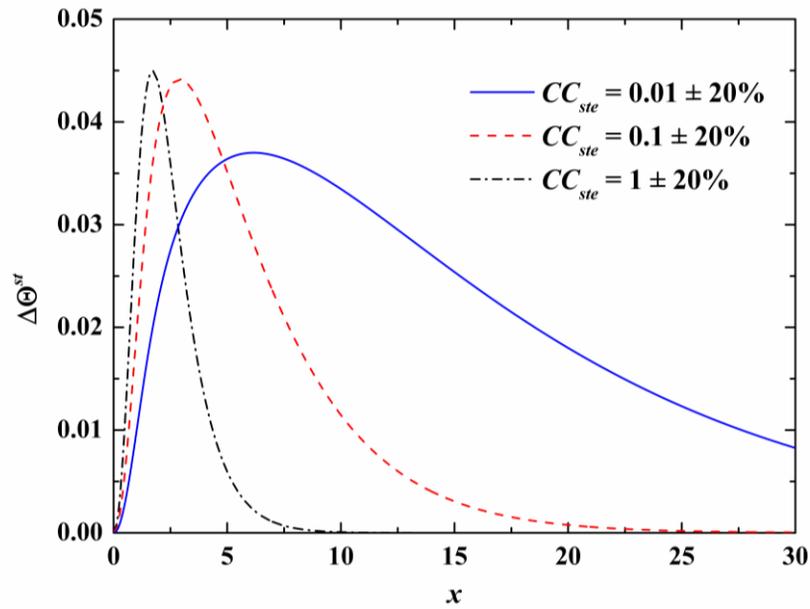

**(b) normalized sensitivities for various $CC_{ste}$**

**Figure 5. (a) Normalized temperature distributions and (b) normalized sensitivities for various $CC_{ste}$.**

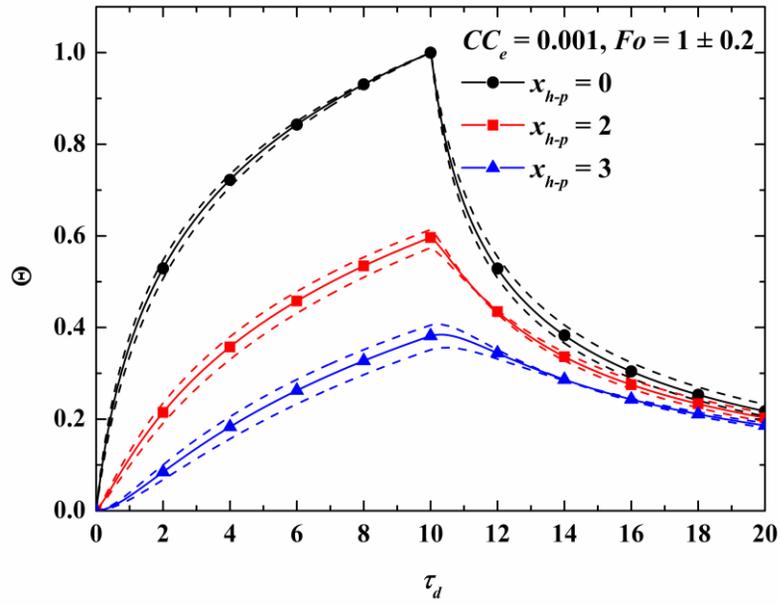

(a) $CC_e = 0.001$

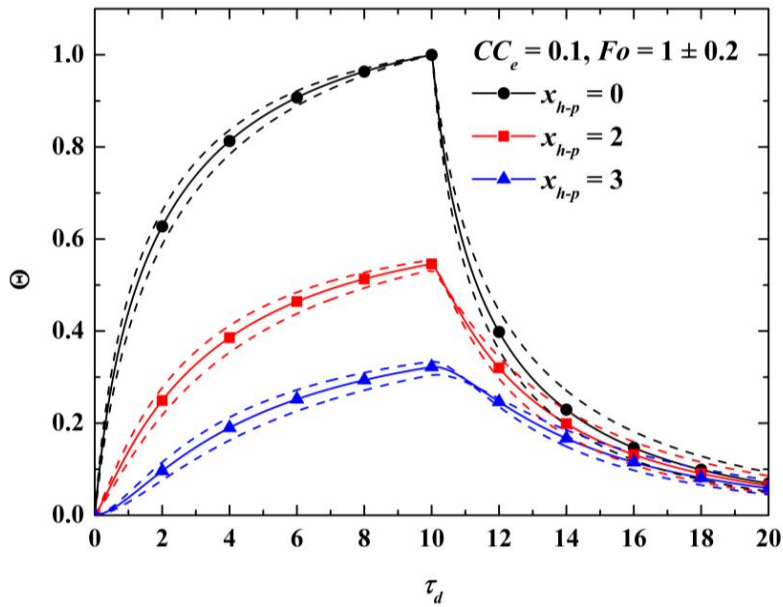

(b) $CC_e = 0.1$

**Figure 6. Normalized temperature variation curves at different positions with *Fo* changes ±20%**

(a) $CC_e = 0.001$; (b) $CC_e = 0.1$.

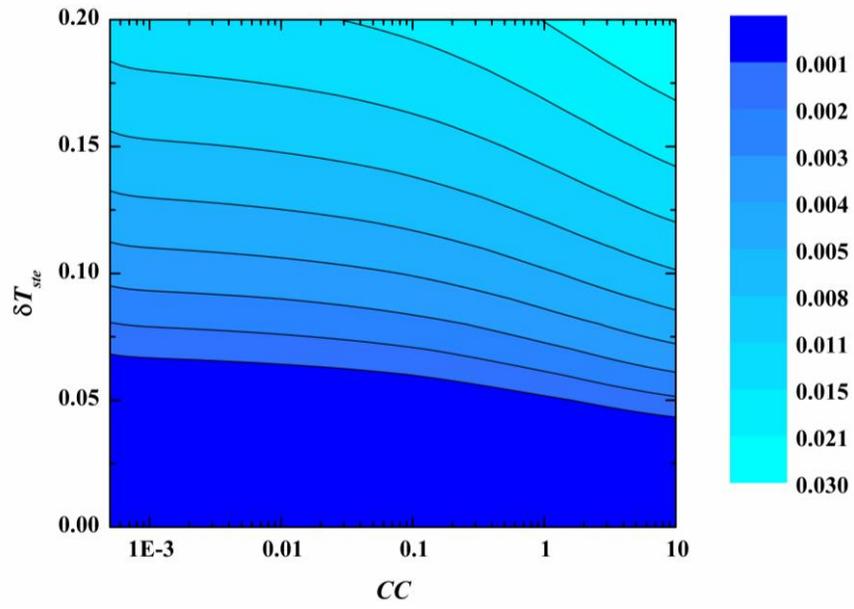

**Figure 7.** System error of *CC* caused by simplifying assumption for various *δT$_{ste}$* and *CC*.

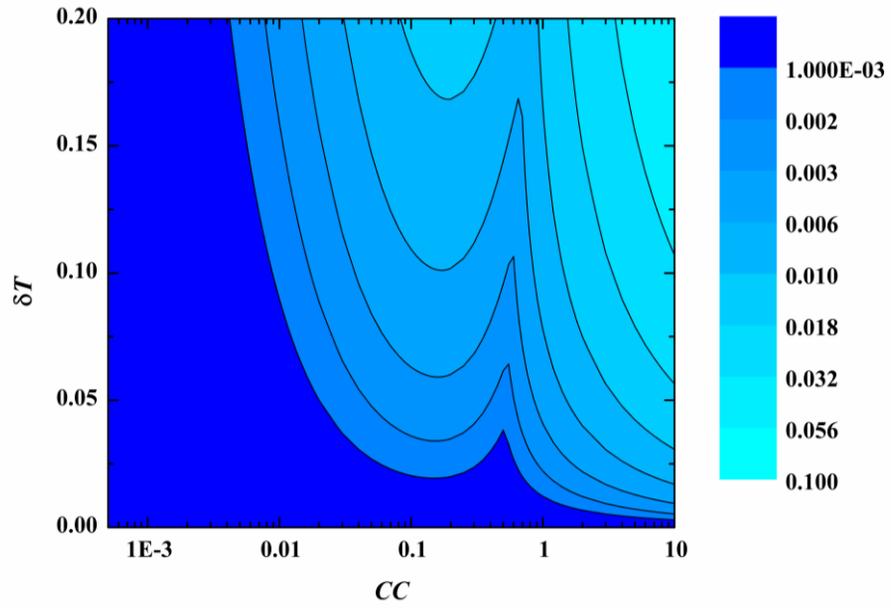

**Figure 8. System error of $α_n$ caused by simplifying assumption for various $δT$ and $CC$.**

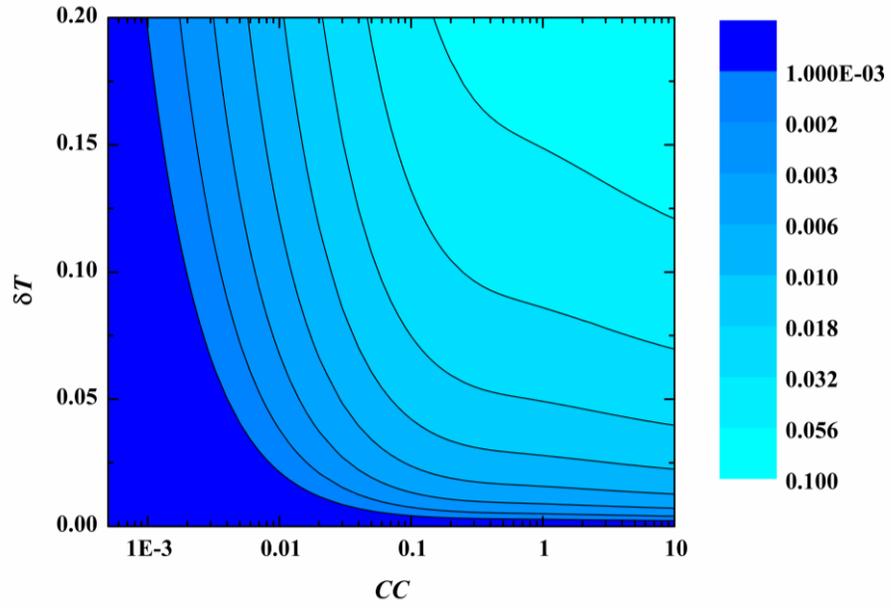

**Figure 9. System error of $α_n$ caused by ignoring substrate temperature rise for various $δT$ and *CC*.**

**Notes on Contributors**

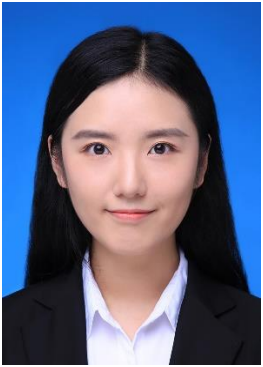
**Aoran Fan** received the B.E. degree from Tsinghua University in 2015. She is a Ph.D. candidate of the Department of Engineering Mechanics at Tsinghua University, China. Her research interests focus on thermophysical properties measurement and interfacial effect at micro/nanoscale.

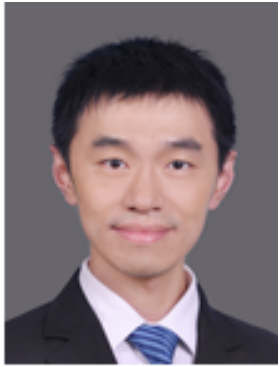
**Haidong Wang** received the B.E. (2006) and Ph.D. (2011) degrees from Tsinghua University. He is now an Associate Professor at Tsinghua University, China. His research interests include nanoscale heat transfer, low-dimensional materials, MEMS technology, and thermal functional devices.

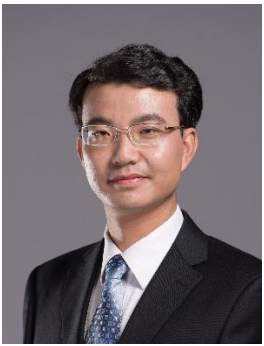
**Weigang Ma** received the B.E. (2006) and Ph.D. (2012) degrees from Tsinghua University. He is an Associate Professor in the Department of Mechanical Engineering, Tsinghua University, China. His current interests mainly focus on micro/nanoscale transport and energy conversion.

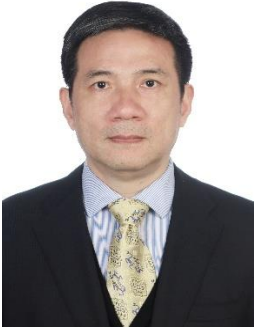 **Xing Zhang** received the B.E. (1982) and M.E. (1985) degrees from Southeast University, and Ph.D. (1988) degree from Tsinghua University. He is a Professor in the Department of Engineering Mechanics, Tsinghua University, China. His current research interests include micro/nanoscale heat transfer, thermophysical properties of nanostructured materials, and efficient use of wind/solar/hydrogen energy.